\documentclass[12pt]{elsarticle}
\usepackage{multirow}

\usepackage{amssymb}
\usepackage{amsthm}

\usepackage[left=2.5cm, right=2.5cm, top=4cm, bottom=3cm, footskip=0.5cm]{geometry}

\journal{ }

\begin{document}

\begin{frontmatter}

\title{Adaptive Multimodal and Multisensory Empathic Technologies for Enhanced Human Communication}
\date{ }

\author[1,2,3]{Roxana Girju}

\address[1]{Department of Linguistics,}
\address[2]{Department of Computer Science,}
\address[3]{Beckman Institute for Advanced Science and Technology, \\ University of Illinois at Urbana-Champaign, \\ Urbana IL 61802}
\address[*]{Corresponding author: girju@illinois.edu}

\begin{abstract}
As digital social platforms and mobile technologies are becoming more prevalent and robust, the use of Artificial Intelligence (AI) in facilitating human communication will grow. This, in turn, will pave the way for the development of intuitive, adaptive, and effective empathic AI interfaces that better address the needs of socially and culturally diverse communities. I believe such developments must consider a principled framework that includes the human perceptual senses in the digital design process right from the start, for a more accurate, as well as a more aesthetic, memorable, and soothing experience. In this position paper, I suggest features, identify some challenges that need to be addressed in the process, and propose some future research directions that I think should be part of the design and implementation.
Such an approach will allow various communities of practice to investigate the areas of intersection between artificial intelligence, on one side, and human communication, perceptual needs and social and cultural values, on the other.\footnote{This position paper was presented at the \emph{Rethinking the Senses: A Workshop on Multisensory Embodied Experiences and Disability Interactions} associated with the ACM CHI Conference on Human Factors in Computing Systems, May 2021.}
\end{abstract}


\end{frontmatter}



\section{Introduction}
The events of recent years have brought both challenges and opportunities to interpersonal communication in all areas of life, especially healthcare. The COVID-19 pandemic, for instance, has taken an enormous toll on people’s mental health. Thus, effective empathic communication has become even more vital in helping people make sense of the unprecedented situation and guiding them through uncertain times. At the same time, the current wave of digital disruption is forcing us to think of new ways to design health encounters and other day-to-day interactions in spaces that accommodate our multifaceted lives, meeting us where we live, work, create, and play. Today we stand at the edge of a revolutionary change especially in the way we understand empathy and compassion in patient-provider interactions, and in the way we teach physicians and other healthcare professionals to experience and express them. 

The emergence of personalized medicine, a paradigm shift from the current “one-size-fits-all” approach, has the potential to balance existing health disparities across diverse populations while improving the quality of healthcare. However, any step toward personalized treatment has to consider social, cultural and environmental factors such as personal factors and preferences, social values, cultural norms, economic and geographic conditions in assessing the cause and progression of disease. Attitudes toward health vary among individuals, families, ethnicities, cultures, and countries, and such factors are increasingly apparent in mental health.  
Specifically, social and cultural groups with core body--mind--soul beliefs and different orientations toward healing practices have experienced first hand the inadequacy of Western medical regimens in identifying and treating mental disorders. 
Several governments have started to address these issues, considering the legal adoption of alternative and complementary medicine like Traditional Chinese Medicine (TCM), Ayurveda, homeopathy, acupuncture, Tai Chi, etc., to improve people’s lives. The most mature integrative health care systems are evident in Asia, in particular in countries like China, South Korea and India who have regulated traditional and complementary medicine into their national health policies. The research community as well has shown increased interest in this direction in a desire to learn how to more effectively engage in such initiatives, as evidenced by a plethora of recent journals, conferences, and symposia on the topic. Some recent memorable examples are briefly introduced below: 

\begin{itemize}
\item \emph{The 3rd Annual Jacalyn Duffin Health and Humanities Conference}, Queen’s University, Canada, January 2021 (https://healthcarehumanities.wordpress.com). Its goals are to create a space where educators and practitioners from various disciplines can discuss relevant issues at the intersections of health, medicine, and the arts and humanities. This year’s conference theme was “Sensations and The Sensational” - stressing the importance of physiological and psychological sensations in health, disease, wellness and illness, as additional clues to one's well-being. The conference is specifically targeting the body -- environment relationship, as the way we perceive our bodies’ sensations is impacted by the world around us and it shapes how we interact with our environment.   

\item 
\emph{The 3rd International E-Symposium on Communication in Health Care: "Advancing Frontiers of Health Communication Research, Education and Practice during the Pandemic"}, Hong Kong, February 2021. The event was part of the Research and Impact Initiative on Communication in Healthcare (HKU RIICH: https://www.hkuriich.org) at the Faculty of Arts, the University of Hong Kong. HKU RIICH is a founding member of the International Consortium for Communication in Health Care (IC4CH). The consortium’s mission is to incorporate cutting-edge communication research into best practice and training for safe and compassionate health care. The consortium is home for other members as well: the Australian National University (Australia), Nanyang Technological University (Singapore), University College London (UK), Lancaster University (UK), and Queensland University of Technology (Australia).

\item 
\emph{The UK’s National Centre for Creative Health Launch Event}, London, UK, 2021. The National Centre for Creative Health (NCCH) was inaugurated on March 9th, 2021. It aims to make creativity integral to health and social care systems. Specifically, their goals are: to address health inequalities; to advance good practice and research; to inform policy; and to promote collaboration. The Centre is working with Integrated Care Systems across the country to explore models for integrating creative health at a systems level.
\end{itemize}

There is growing evidence showing that an integrative approach to healthcare has the potential to provide better interpersonal care to patients \cite{Pun2019DeliveryOP,MTM2020,TIM-NA2020,NI20201149} by engaging healthcare modalities along sensorial, cognitive, and emotional dimensions for a diversified personal experience. 
Western medicine comes with its own history, language, knowledge practices, regulations, expectations, and scientific evidence to address healthcare concerns, often targeting one area of health and healing. Complementary and alternative practices, on the other hand, bring in a holistic perspective of the patient’s body and the environment in which they work and live. 
In both diagnosis and treatment, all of the senses (touch, sight, hearing, taste and smell, proprioception, etc.) and psychological resources of the patient are considered. 
This allows the practitioner to focus on the patient's internal processes that support the overall goal of healing as well as the patient's embedded interaction with their environment, challenging the patients' negative state of mind, and addressing their unhealthy feelings and thoughts. 
Unlike Western medicine’s tendency to target one area of healing, these methods focus on the highly interconnected physical, mental and spiritual dimensions of well-being. 

Given the global increase in preferences towards alternative health practices \cite{GrandViewRes}, collaboration among different medical and knowledge systems becomes even more crucial. 
It has become more evident that modern and traditional medical systems are potentially complementary rather than antagonistic. Thus, ethnic and 
cultural considerations can and should be integrated further into the modern health delivery system to improve care and health outcomes. Each culture has certain core beliefs about the body, mind, and soul as well as about health and well-being. Hence, the Western concepts may not be always and directly applicable in all socio-cultural scenarios.

In this paper I suggest some important aspects of smart interface design that could enhance human interaction capabilities online and inform future empathic communication technologies in healthcare and other fields that can benefit from such a communication framework.

\begin{flushleft}
\textbf{Alternative Healing Practices and the Senses}
\end{flushleft}
We use all our senses of sight, hearing, touch, smell, taste, proprioception, etc. to make sense of the world around us. Just as the quantity and quality of food we eat impact our health and shape our body, our sensory impressions contribute to the quality of our thoughts, sensations, and emotions, affecting our nervous system. Everyday life experiences provide us with a wide variety of sensory perceptions, where each sense organ can be overused, underused or misused, affecting our emotional, physical and spiritual health. Although many people can live well and safe with sensory impairment, any disturbances or loss of our senses can have a profound impact upon us. Many conventional healthcare practices like Ayurveda use a variety of approaches to integrate holistic sensory therapy into healing the body and mind. In Ayurveda, there are three main types of mind--body constitutional energy known as Doshas: Vata, Pitta and Kapha.  When the Doshas are combined with knowledge of the elements and the senses, practitioners can identify patterns of imbalance that can lead to a variety of ailments. In Ayurveda, also known as the “Science of Life”, all things in nature are made up of a combination of five elements: air, water, ether, fire and earth. Human beings are also seen to be a sum of these five elements -- a uniquely balanced blueprint. Understanding the interplay among the elements for each individual will determine one's personality: who they are as a person, how best to understand them, and how they should be treated. 

Ayurveda's central aim is to find harmony and balance among all of these five elements and three doshas. Imbalances in each of the doshas and elements can indicate disturbance patterns of these senses. Ayurveda can foster a return to balance in an aesthetically pleasing and comforting practice associated with positive memories and sensations triggered by smells, tastes, touch, visual and audio cues \cite{Halliburton2003}. Just as certain events, objects, and spaces can trigger trauma, so various healing modalities can evoke positive memories and emotional states. \cite{Schreuder-etal2016}.

\section{Conceptual Design: The Sensorium of Empathic Communication Interfaces}
Every human interaction is a sensory experience, a complex symphony of sensory perception guiding our thoughts, emotions, and interactions. Yet, current digital design practices of (empathic) communication tend to focus merely on sight \cite{FRIESEM2016}. Moreover, designing AI-powered interfaces that support situated multimodal and multisensory empathic interactions requires effective functioning and coordination among a number of subsystems, from complex Artificial Intelligence components (speech, static image, video, and natural language processing) to interface design, to the nature of interactions themselves that have to consider all relevant aspects of being human. As telehealth, and virtual communication in general become more prevalent, the use of AI in facilitating human interaction will grow, pushing the need for intuitive, adaptive, and efficient emotionally-intelligent AI interfaces.
Trained on a large number of patient - provider interactions, for instance, the AI system can learn complex patterns from a bi-directional process of meaning negotiation based on a mixture of verbal and social cues, like gaze, gestures, posture and head movements.
Such patterns are then tightly correlated with specific emotions and needs. The more the system learns, the better it becomes at figuring out the emotional and mental state of the patient, thus being able to guide the dialogue in real time and later teach the provider where he/she could have done better suggesting more empathic styles of communication. But such interactions are not only about \emph{what} to say (or not to say) and \emph{how} to say it. In conversation, we pay attention to everything, from the pale virtual background colors to the soft sweater the interlocutor is wearing, to the smell memories the candle behind her evoke. 

In this position paper I argue that such smart empathic interfaces should be designed and implemented as interactive environments that facilitate a versatile multimodal and multisensory engagement for a more efficient as well as a more aesthetic, memorable, and soothing experience. This approach allows us to investigate the areas of intersection between artificial intelligence, on one side, and human communication, perceptual needs and social and cultural values, on the other.
For this, we need a principled framework that includes not only the main modalities (vision, speech, linguistic content), but also the senses in the digital design process right from the start. I outline below some features, challenges, and future research directions that I think should be considered in the design and implementation of such a framework.

\subsection{Accessing the Patient’s Personal and Cultural Profile and Adapting the Interface to their Needs}

In Ayurveda, Doshas represent the three humors or forces of the body, which bring health when in balance, and produce diseases when out of balance. 
Each Dosha determines to some extent one's body shape, food preferences and digestion type, as well as their thoughts, emotions, and behavior.
A Vata personality, for instance, is known to be naturally drawn toward colorful sights, bright colors, scary movies, and varied vocal tones. Vata disturbances show symptoms of anxiety, fear, pain hypersensitivity, and insomnia. Restorative sensory therapies can address and improve the patient’s healing capacity through calming sights (i.e., trees, sunset), warming colors (i.e., gold, orange, green), sweet and sour scents, and slow-paced and relaxing sounds. A Pitta energy, on the other hand, is more fiery and when disturbed, shows more irritability, anger, and judgment. Remedies include soothing sights (i.e., lakes, green landscapes), cooling colors (blue, white, pastels), sweet scents, and nature sounds. 
Ayurveda practitioners learn to use their own senses to diagnose aspects of an individual’s constitution and humoral imbalance. Such a detailed personal profile of the patient can better inform both a medical diagnosis as well as the interface design in how to appropriately use the senses for each patient's personality type and constitution in technology mediated patient -- provider interactions. This kind of interfaces should be adaptive, designed to be constantly evolving and generating relevant experiences, based on the patient’s personality profile, life experience and needs.

\subsection{Creating Sensory Experiences for the Exploration of Self and Other}

In such multimodal, multisensory interface-mediated interactions, one can experience various types of shifts in sensational states - e.g., how we experience the level of pain, or changes in taste or touch. The complex interface should be designed to directly or indirectly stimulate and monitor such changes for a better understanding of self and other.

\subsection{Exploring Synesthetic Interactions}

Senses are highly interconnected and could lead to synesthetic experiences. Synesthesia is a Greek term -- \emph{syn} (together) and \emph{anesthesia} (sensation) -- where the senses are intertwined or come together in some way: 
colors are tasted; images evoke odors; sound can stimulate touch, etc. This phenomenon has long been used in marketing strategies to bring to light cross-sensory connections \cite{Velasco-Orbist2020} 
by employing digital design elements to activate modalities beyond sight and hearing.
Given one's age, gender, ethnic background or local climate, certain sensory inputs tend to elicit particular reactions and affect how we feel. However, as in any kind of design practice, some design decisions may have undesired sensory consequences.
For example, in some scenarios and especially for particular individuals, the color red has been shown to be too stimulating, potentially raising blood pressure and heart rate.
Other modalities and cross-modal, cross-sensory simulations may make one feel tired, dizzy, or cranky. 
Although these modalities are highly intertwined, some can and do appear as distinct and salient in their own right. 
It is thus important to study thoroughly sensory combinations and their effects on the  interaction and its participants.

Empirical research practices have to be ethically applied to discern how sensory features help or hinder the user experience in various situated scenarios.
Moreover, researchers and practitioners should be aware of the fact that combining multiple modality and sensory elements might lead to sensory overload. Depending on the situated context and the interaction participants, sensory elements should be subtle, working in the background to contribute to an empathic interaction, and not to hinder it. Isolating the individual senses and then considering the various sense combination latices and studying the sensory and care implications of such design elements are necessary steps toward a better understanding of the complex human -- human and human -- environment embodiment. 



\subsection{Understanding the User's Motivation, Knowledge and Culture and Considering the Practitioners' Own Biases and Limitations}

Most alternative healthcare practices take into consideration the time, place, climate, diet, the type of work patients perform, their social and cultural values, their hopes, fears and concerns.
Thus, any communication with the patient requires an understanding of such factors in addition to patients' preferences.  
The patient's narrative plays an important part in empathic communication informing on the complexities of their lives which can affect their well-being.
Medical practitioners need to be aware and sensitive to their patients' condition and to take into account a much wider and complex context in assessing symptoms and suggesting treatments. But most importantly, practitioners have to be aware of their own biases and limitations, and to constantly try to find ways to better understand socially and culturally different populations. This will allow them to be more empathetic and more open to understanding and trust. Smart empathic interfaces coupled with active partnerships of various healthcare practices should facilitate the application of theories, examination, diagnosis and treatment methods toward a better understanding of the patients’ physiological, psychological, social and cultural needs. This is particularly important, given the major mental health issues rampant in the aftermath of the pandemic, especially in marginalized and underserved communities.


\vspace{0.05in}
\begin{flushleft}
\textbf{Discussion Points}
\end{flushleft}

In this position paper and workshop discussion I am particularly interested in exploring issues related to the design of human empathic interaction interfaces in professional and private life, with a special interest in healthcare. Specifically, I will address the following questions: 
\begin{enumerate}
\item Sensory compensation refers to the lack or alteration of one sensory modality that could change the distribution of input from the other sensory modalities.
How can we best capture this cooperation of human sensory modalities to inform the interface design?
It would also be interesting to explore (in collaboration with neuroscientists) the amount of time and effort required for the human nervous system to accommodate new inputs from various individual sensory receptors.

\item Various challenging life situations tend to trigger intense negative emotional states which are often later re-experienced as a series of embodied sensations. How does the particular cultural or personal valence of these states impact the intensity level at which such sensations are re-lived?

\item How does the growing interest in integrating modern and traditional healing practices and forms of diagnosis and treatment influence how senses and bodily sensations are framed and perceived by different segments of the population and cultural communities?
\end{enumerate}

\section{Conclusion}

Adaptive multimodal and multisensory empathic interfaces can play a very important role in creating perceptual spaces where people can experience interactions in new, accessible,  inclusive, and ethical ways. 
In this paper, I suggest that designers and developers of such smart empathic interfaces should pay particular attention to sensory exploration allowing us to interpret and reinterpret our senses, and combine them in different ways for a more effective communication. With the user in mind, such a framework can help us explore how senses can inform the future of AI-powered empathic technologies. 
%

The Italian philosopher, Emanuele Coccia \cite{Coccia2016} once said “we consider ourselves as rational, thinking, speaking beings — yet, for us to live is above all to look, taste, touch, or smell the world. Our primary relationship with the world as it exists is not an act of immaterial contemplation, nor a practical, moral, and ethical fact. Our relationship with the world is a sensible life: sensations, smells, images, and above all, a restless production of sensible realities.”

I hope this paper will facilitate the creation of a space and community that cultivate inclusion and creative expression in empathic design informed by and for a more immersive human experience. I also hope the issues and suggestions addressed here open up the necessary dialogue that brings together medical professionals from different practices with researchers in human-computer interaction, artificial intelligence, natural language processing, as well as humanities, social sciences and the arts.
We need to build systems that guide both the practitioner and the patient in challenging healthcare and life situations taking advantage of a mix of interconnected system components and practices to make digital interactions more useful, efficient, and memorable. 


As Carlos Velasco and Marianna Orbist's call for action \cite{Velasco-Orbist2020} regarding the implications of multisensory experiences for individuals and society, I, too, want to stress here an aspect of considerable importance for healthcare communication in virtual spaces. This is how medical professionals with different practices should work together and engage in referral to deal with different aspects of the healthcare process and provide a better, safer and more soothing healing experience to the patient. In North America, in particular, we know little about practices and patterns of referral and collaboration between traditional biomedical experts and complementary and alternative medicine practitioners. This is a rather puzzling and concerning state of affairs of our healthcare system, given that complementary and alternative medicine has become more and more popular among various population segments, and more and more practitioners are becoming interested in integrated medicine. \cite{GrandViewRes,frass2012use,harris2012prevalence,wiles2001gentle}.





\vspace{0.15in}
\textbf{Short Bio}

Roxana Girju is professor of computational linguistics at the University of Illinois at Urbana-Champaign, where she studies, designs, and builds empathic language communication technologies. She is also involved in initiatives that raise awareness about the need for a more strategic approach to account for human values in artificial intelligence. Dr. Girju is a longtime proponent and user of alternative and complementary medicine techniques of preventive care.
She brings in an expertise of over twenty years in the corporate world and academia as Professor of Linguistics and Computer Science (having worked at Baylor University, Northwestern University, and the University of Illinois at Urbana-Champaign). Located at the intersection of language, technology, and society, her research interests center around various aspects of language that can inform frictionless, successful human communication technologies, with application to  healthcare, social media, smart communication interfaces, and education. This position paper and research presentation are part of a larger project that looks at how human sensory experiences emerge through language to foster empathy and minimize social distance in both physical and virtual environments.
Dr. Girju hosts her own podcast, the \emph{Creative Language Technologies}, which explores the multifaceted aspects of this emerging field, at the intersection of Science, Technology, Engineering, Math and Medicine (STEMM) with the broader sector of Humanities, Social Sciences, Arts and Culture (HSSAC). The podcast aims to revitalize technological imagination and promote creative and diverse themes with social impact. It can be accessed on any major podcast platform as well as on the web (https://player.fm/series/creative-language-technologies).

\bibliographystyle{elsarticle-num-names} 
\bibliography{cas-refs}





\end{document}